# Design and Analysis of Plasmonic-Nanorod-Enhanced Lead-Free Inorganic Perovskite/Silicon Heterojunction Tandem Solar Cell Exceeding the Shockley-Queisser Limit

Md. Sad Abdullah Sami[1], Arpan Sur[1] and Ehsanur Rahman[1]*

[1] Department of Electrical and Electronic Engineering, Bangladesh University of Engineering and Technology, Dhaka, 1000, Bangladesh

*Email: ehsaneee@eee.buet.ac.bd

## Abstract

The pursuit of sustainable and highly efficient energy conversion necessitates a transition from toxic and unstable materials to environmentally friendly alternatives. This work presents a simulation-based numerical investigation of a fully inorganic, lead-free tandem solar cell that employs cesium tin-germanium tri-iodide ($CsSnGeI_3$) as the top cell absorber and crystalline silicon (c-Si) as the bottom cell absorber, configured in a silicon heterojunction (SHJ) arrangement. Utilizing $CsSnGeI_3$ as a lead-free perovskite presents a promising solution to the toxicity concerns associated with conventional lead-based perovskites. To further increase near-infrared absorption and reduce the required thickness of the c-Si layer, an ultra-thin gallium antimonide auxiliary absorber is integrated into the SHJ bottom cell. Optical and electrical simulations, conducted using finite-difference time-domain and drift-diffusion modelling, demonstrate that the optimized tandem structure attains a power conversion efficiency of 34.93%, surpassing the Shockley-Queisser limit established for single-junction Si cells. Furthermore, the optimized device showcases an open-circuit voltage of 1.93 V, a short-circuit current density of 21.30 mA/cm$^2$, and a fill factor of 84.74%. Performance is additionally enhanced by incorporating cylindrical gold nanorods within a $Si_3N_4$ dielectric medium positioned at the rear of the bottom cell, thus amplifying light absorption through plasmonic effects. Notably, the tandem cell sustains high efficiency even without the plasmonic structure, thereby providing flexibility for cost-effective fabrication. This work underscores the viability of all-inorganic, lead-free tandem cells for next-generation photovoltaics, guided by simulated results that pave the way for high-efficiency, non-toxic solar energy solutions and further experimental validation.

# Keywords

**Shockley-Queisser limit, Perovskite solar cell, Lead-free, Inorganic, Silicon heterojunction, Tandem solar cell, Plasmonic nanorods**

# 1. Introduction

The world population has increased exponentially since the 19th century, reaching around 8 billion in 2022, and is projected to exceed 9 billion by 2040 [1]. Due to this rise in population and lifestyle changes, energy demand is increasing daily. Global energy demand is expected to surpass 35,000 terawatt hours (TWh) by 2040 [2]. The energy produced today remains predominantly dependent on fossil fuels, but the availability of fossil fuel sources is expected to diminish around 2112 [3]. A possible solution to this issue is the use of renewable energy. Among renewable energy sources, solar energy has the most potential [4]. Historically, semiconductor materials have played a crucial role in photovoltaic (PV) cells, with silicon (Si) being utilized in PV cells for about 70 years. Due to this extended development period, Si-based PV cells dominate the current solar cell industry, accounting for 97% of total PV cell production in 2023 [5]. Si is abundant, non-toxic, stable, and has a bandgap of 1.12 eV, which makes it a suitable candidate for absorbing a wide range of the solar spectrum. The fabricated crystalline Si (c-Si) single-junction solar cell has achieved a remarkable power conversion efficiency (PCE) of 27.09% [6]. The maximum possible PCE for a single-junction Si solar cell is restricted to 29.43% [7]. Si has an indirect bandgap, which contributes to its low photocurrent generation. Its thickness must be significantly increased to produce a sufficient photocurrent with the current industry standard for c-Si absorber layer thickness at 180 μm [8].

Single-junction solar cells are constrained by the Shockley-Queisser (SQ) limit, which defines the maximum theoretical efficiency achievable by a single p-n junction photovoltaic device under standard illumination conditions [9]. This limit is approximately 33% under standard AM 1.5G solar spectrum conditions. Over the years, different approaches have been taken to increase the PCE of Si PV cells. Some prominent ones include passivated emitter and rear contact (PERC) cells and Si heterojunction (SHJ) solar cells. PERC solar cells have achieved a maximum recorded efficiency of 24.5%, which was produced in 2022 by Trina Solar [10]. In PERC solar cells, the rear side is passivated so that unabsorbed light can be reflected from the back of the cell. The top layer of a PERC solar cell is textured to enhance the absorption of incident light [11]. In 2023, LONGi achieved a significant advancement in Si solar cell technology by developing a solar cell with an efficiency of 27.09%, utilizing a Si heterojunction (SHJ) [6]. Here, a thin layer of a-Si was used with a 110 μm thick c-Si layer to produce a SHJ solar cell. Thus, a higher PCE was achieved. However, this efficiency remains below the theoretical limit, prompting the exploration of various strategies to further enhance the PCE of Si solar cells. Using a multi-junction tandem structure is a prominent one among them. In Si-based tandem cells, a suitable top cell is chosen, and Si is kept as the absorber layer for the bottom cell. The tandem cell's top cell absorber layer has a bandgap typically around 1.5 to 1.9 eV. This ensures that both the top and bottom cells can absorb an adequate solar spectrum, which is crucial for the tandem structure to function correctly [12].

Various materials have been explored as potential top cell absorbers in Si-based tandem solar cells. Among them, group III-V materials [13], kesterites [14], and perovskites [15], have received considerable attention. Due to their direct bandgap, these materials exhibit strong absorption and photocurrent generation capabilities. Specifically, the perovskite crystals demonstrate unique properties, including a tunable bandgap, low Auger recombination loss, high carrier mobility, high absorption coefficient, and a long charge diffusion length [15][16]. The general chemical formula of perovskite compounds is $ABX_3$, where A typically denotes a larger organic or inorganic cation, B represents a smaller metallic cation, and X corresponds to an anion, commonly a halide [17], [18]. Moreover, perovskite materials feature a tunable bandgap of 1.3 to 2.2 eV, allowing seamless integration into tandem structures for targeted solar spectrum absorption. Additionally, perovskites exhibit greater resistance to crystallographic defects than conventional semiconductor materials [19][20]. Perovskite-on-Si tandem solar cells have recently garnered significant attention within the PV industry. As of September 2024, LONGi has reportedly fabricated a perovskite/Si tandem solar cell with an impressive efficiency of 34.85% [21]. The quest for an optimal perovskite material that can effectively complement existing Si technology in tandem architectures remains ongoing. While a variety of perovskite compositions were initially proposed, lead-based perovskites have emerged as the primary focus of continued investigation, with methylammonium lead iodide ($MAPbI_3$) being the most extensively studied [22][23][24][25], [26]. It possesses a direct bandgap of 1.55 eV, enabling it to absorb a significant portion of the solar spectrum with a thickness of only a few hundred nanometers, compared to 100-200 micrometers in the case of Si. It also has a lower fabrication cost compared to Si [27], making it an industry favourite. Moreover, it is efficient for transporting positive and negative charge carriers over long distances. Under the AM 1.5G standard solar spectrum, the electron and hole diffusion lengths have been shown to exceed 175 μm [16]. Thus, MAPbI$_3$ can help build a tandem cell that can potentially achieve a PCE beyond 41% [23].

Despite their numerous advantages, lead-based perovskites present significant drawbacks, particularly concerning environmental and health-related issues. The inherent toxicity of lead compounds poses a major obstacle to the long-term sustainability of lead-based perovskite solar cells [28][29][30]. Lead ions have the potential to contaminate soil and water resources, resulting in adverse effects on human health, as well as on animal and plant life [31][32]. For instance, MAPbI$_3$, commonly used as the light-absorbing layer in perovskite solar cells, typically has a thickness of around 500 nm, which corresponds to a lead content of approximately 0.6 g/m² [33][34]. Consequently, the development of lead-free alternatives has become a critical research priority for advancing the commercial viability and environmental safety of perovskite solar technologies. Lead ion in the perovskite structure has been replaced by alternative ions such as $Sn^{2+}$, $Cu^{2+}$, $Bi^{3+}$, $Mn^{2+}$, and $Ge^{2+}$, which are less toxic than $Pb^{2+}$, and could be utilized to form a lead-free perovskite structure. Tin-based perovskites, for instance, offer benefits like reduced toxicity, enhanced stability, and improved environmental compatibility [18], [35], [36]. Chen et al. [37] has shown that partial substitution of Sn(II) with Ge(II) in CsSnI$_3$, forming the mixed-cation perovskite CsSn$_{0.5}$Ge$_{0.5}$I$_3$, significantly enhances film stability. This improvement in structural robustness is attributed to the favourable Goldschmidt tolerance factor (0.94) and octahedral factor (0.4) associated with the alloy. Additionally, the high oxidative reactivity of Ge(II) facilitates the rapid formation of a uniform, ultrathin (<5 nm) native oxide layer on the surface, effectively passivating the material and conferring superior environmental stability compared to the benchmark MAPbI$_3$ perovskite. These characteristics not only enhance film-level chemical stability but also indicate improved device-level operational reliability under exposure to heat, moisture, and light stressors. Therefore, CsSnGeI$_3$ is a promising candidate for stable long-term photovoltaic performance, especially when combined with inorganic charge transport layers like TiO$_2$ and Cu$_2$O, which are also known for their chemical durability [38], [39].

Due to these advantages, CsSnGeI$_3$ has recently been explored as a lead-free alternative to MAPbI$_3$. Moreover, CsSnGeI$_3$ has a similar bandgap to MAPbI$_3$, making it easy to replace MAPbI$_3$ in a tandem-based structure with CsSnGeI$_3$ as the absorber layer [40][41][42][43]. Lead-based organic perovskites are typically used in perovskite/Si tandem cells to ensure high efficiency [44], [45]. At times, these lead-based organic perovskite/Si tandem cells can also surpass the SQ limit [45]. Sarker et al. [46] has exceeded the SQ limit by using a lead-free perovskite/Si tandem cell. Moreover, the perovskite used in that study was an organic one, which can potentially cause instability in the structure. However, utilizing both lead-free and inorganic perovskite materials in a perovskite/Si tandem structure typically results in a lower PCE compared to their lead-based organic counterparts [42], [47]. Moreover, previously reported perovskite/Si tandem structures neglected the issue of significant c-Si layer thickness (usually exceeding 100 μm) [45][46][44], [48].

To address these challenges, this study has thoroughly investigated the CsSnGeI$_3$/Si tandem solar structure. This study has achieved a PCE beyond the SQ limit comparable to lead-based organic perovskite/Si tandem cells by using a lead-free inorganic perovskite. Moreover, this study has accomplished this while also minimizing the thickness of the c-Si absorber layer (2 μm). Gallium antimonide (GaSb) is added to the bottom SHJ cell as an auxiliary absorber layer, improving absorption specially in the near-infrared (NIR) range [49]. Furthermore, plasmonic gold (Au) nanorods are incorporated at the rear side of the tandem cell to improve its performance [50], [51], [52]. Although the incorporation of nanorods on the rear side of solar cells has been previously investigated, their integration into a CsSnGeI$_3$/Si tandem architecture has yet to be reported. Consequently, this study represents the inaugural demonstration of this approach, with the objective of enhancing light trapping and overall device performance within lead-free tandem configurations. Initially, optimization of the thicknesses of different layers has been conducted. The thicknesses of CsSnGeI$_3$ and c-Si absorber layers, along with the p-doped and n-doped a-Si layers, titanium dioxide (TiO$_2$) and cuprous oxide (Cu$_2$O) transport layers, have been optimized by analyzing generated contour plots of PCE, open-circuit voltage ($V_{oc}$), short-circuit current density ($J_{sc}$), and fill factor (FF). Parametric sweeps have also been performed to determine the optimal thicknesses of the anti-reflecting coating (ARC), GaSb, and rear passivation layers. This study further explores an advanced tandem solar cell architecture by incorporating cylindrical-shaped Au nanoparticles embedded in a dielectric medium at the rear side of the bottom silicon cell. A comprehensive investigation was conducted to determine the optimal dielectric material, nanoparticle geometry, and particle radius to enhance light management and improve device performance. In addition, the doping concentrations of all layers within the tandem structure were systematically optimized. The combined effect of these enhancements led to a significant improvement in the PCE of the tandem cell, enabling it to exceed the SQ efficiency limit.

## 2. Device Structure and Simulation Methodology

## 2.1. Device Structure

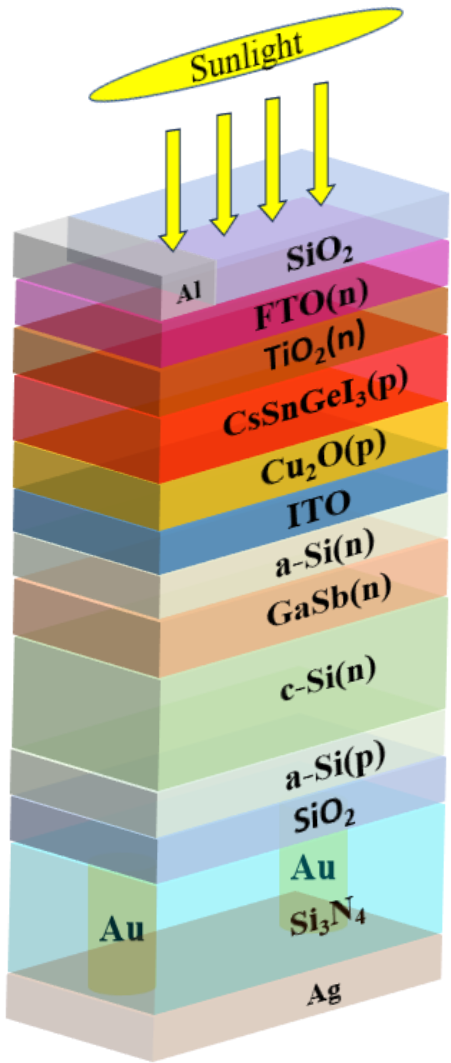 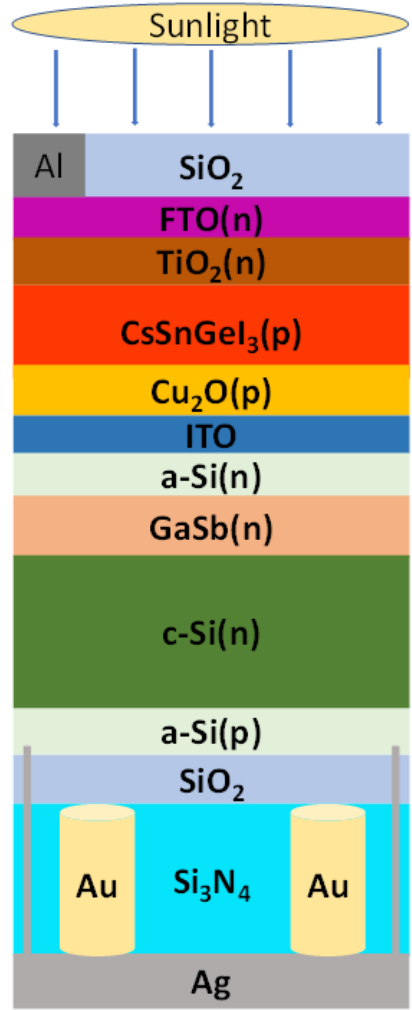

**Figure 1:** Cross-sectional schematic of (a) 3D and (b) 2D views of the CsSnGeI$_3$/Si tandem solar cell incorporating a plasmonic back reflector, where cylindrical Au nanorods are embedded in Si$_3$N$_4$ dielectric medium. An ITO recombination layer connects the perovskite top cell and the SHJ bottom cell. The SiO$_2$ layer on the top serves as an anti-reflective coating (ARC) layer, while the rear SiO$_2$ layer functions as a passivation layer.

Fig. 1 shows the tandem structure used in this study, where CsSnGeI$_3$ and Si are the primary absorbers of the top and bottom cells, respectively. At the top of the structure, there is an ARC layer made of SiO$_2$. Aluminium (Al) serves as the finger electrode for the tandem solar cell. The top cell features a perovskite absorber with the following layer configuration: FTO/TiO$_2$/CsSnGeI$_3$/Cu$_2$O. Fluorine-doped Tin Oxide (FTO), a transparent conductive oxide (TCO) layer, serves as the front electrode in this tandem structure, enabling light transmission and efficient electron collection [53]. The electron transport layer (ETL) consists of a thin layer of TiO$_2$, while the hole transport layer (HTL) is made of Cu$_2$O. The perovskite absorber layer is positioned between the ETL and the HTL layers. The ETL and HTL materials were selected based on the calculation and comparison of the conduction band offset (CBO) and valence band offset (VBO) of various materials [45].

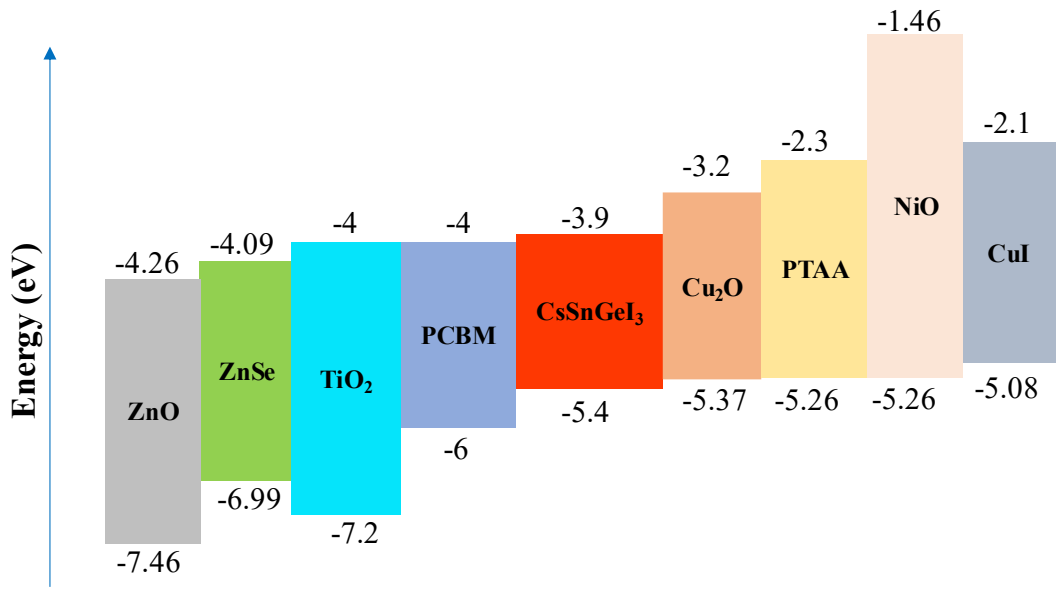

**Figure 2:** Energy level diagram of CsSnGeI$_3$ with various ETLs and HTLs.

In fig. 2, some ETLs and HTLs were compared. The hole transport layers (HTLs) include Cu$_2$O, poly[bis(4-phenyl)(2,4,6-trimethylphenyl)amine] (PTAA), nickel(II) oxide (NiO), and cuprous iodide (CuI), with their valence band offsets being -0.03 eV, -0.14 eV, -0.14 eV, and -0.32 eV, respectively. Materials with a highly negative VBO in the HTL are undesirable because they increase the likelihood of interface recombination, which can lead to a reduction in V$_{oc}$ [38]. Additionally, inorganic materials typically exhibit higher hole mobility than organic HTL materials, resulting in higher J$_{sc}$ [54]. Cu$_2$O was selected as the HTL layer in the top cell due to its lowest VBO and its classification as an inorganic HTL. The ETLs illustrated in fig. 2 include zinc oxide (ZnO), zinc selenide (ZnSe), TiO$_2$, and [6,6]-phenyl-C$_{61}$-butyric acid methyl ester (PCBM), with their conduction band offsets being -0.36 eV, -0.19 eV, -0.1 eV, and -0.1 eV, respectively. A negative CBO is suitable for ETLs, as a positive CBO creates a barrier to the flow of photo-generated electrons [55]. An ETL with a low negative CBO is advantageous, resulting in a higher V$_{oc}$. Among the ETLs depicted in fig. 2, TiO$_2$ and PCBM have the lowest negative CBO of -0.1 eV. However, since PCBM is an organic ETL and organic ETLs generally have lower electron mobility than inorganic ones [56], TiO$_2$ is chosen as the ETL for the top cell. An Indium Tin Oxide (ITO) layer connects the top and bottom cells by serving as a recombination layer. This layer is important to ensure continuous current flow in the tandem structure as without it, carriers would build up at the interface, and current continuity between top and bottom cells would break[57]. The bottom cell features a SHJ structure, utilizing two ultra-thin amorphous Si (a-Si) layers. The n-doped a-Si layer functions as the emitter, while the p-doped a-Si layer acts as the back surface field. Additionally, a thin layer of GaSb is used as an auxiliary absorber layer in the bottom cell, alongside a c-Si layer. On the rear side of the bottom cell, a SiO$_2$ layer is employed as the rear passivation layer. Beneath it lies a Si$_3$N$_4$ dielectric layer, which contains an array of cylindrical Au nanorods with a periodicity of 750 nm. At the very bottom of the structure, a Ag mirror functions as both a reflective layer and a back contact. Small contact windows are opened on the Si$_3$N$_4$ and SiO$_2$ layers to connect the Ag back contact with the p-doped a-Si layer.

## 2.2. Proposed Fabrication Process:

Ag contact can be created by screen-printing Ag paste for the back contact, then co-firing it in a belt furnace. A thermal SiO$_2$ layer can be grown for surface passivation before metallization, and a Si$_3$N$_4$ dielectric layer can be applied using the Plasma Enhanced Chemical Vapor Deposition (PECVD) method [58]. Laser openings can expose the a-Si layer locally through the dielectric and rear passivation layers. This setup improves carrier selectivity and internal reflectance, which supports the high efficiency of PERC devices. Au cylindrical nanorods can be produced by refining the seed-mediated synthesis of gold nanorods [59], which can then be embedded in the dielectric layer via a wet-coating process [52]. Next, a thin p-type a-Si layer can be deposited through PECVD above the rear passivation layer. Following this, a high-quality c-Si layer can be bonded or integrated epitaxially on top of it [60], [61]. A GaSb layer can then be grown on the c-Si using

Molecular Beam Epitaxy (MBE), beginning with an AlSb nucleation layer at about 450 °C [62], [63], [64]. An n-type a-Si layer is deposited above GaSb using low-temperature PECVD. ITO on a-Si can be created through a two-step radio frequency (RF) sputtering method [65]. At first, a low-damage ITO seed layer can be applied at 50 W to minimize interface degradation, followed by a high-power deposition at 100 W to enhance conductivity and transparency. $Cu_2O$ can be formed on ITO substrates via electro-deposition using an alkaline solution containing $Cu^{2+}$ ions and a chelating agent like citric acid [39]. $CsSnGeI_3$ can be created on $Cu_2O$ by spin-coating a precursor of $CsSnGeI_3$ onto the $Cu_2O$ layer, followed by a carefully controlled annealing process [66], [67]. $TiO_2$ thin films can be deposited via a sol-gel spin-coating technique, enabling uniform coverage with controlled thickness [68]. FTO thin films can be deposited over the $TiO_2$ layer via RF magnetron sputtering at room temperature [69]. Then, a $SiO_2$ antireflection coating layer can be fabricated through a Physical Vapor Deposition (PVD) technique, which is based on the thermal decomposition of polydimethylsiloxane (PDMS) [70]. Finally, Al can be deposited via electron beam evaporation under high vacuum conditions, with controlled deposition rates [71].

## 2.3. Research Methodology

The simulation consists of two parts, specifically optical and electrical simulations. Optical simulations were utilized to compute the carrier generation rate. This data served as the input for subsequent electrical simulations, from which the characteristic PV parameters, such as PCE, $J_{sc}$, $V_{oc}$, and FF, were extracted. In the optical simulations, the refractive index (n) and extinction coefficient (k) values as a function of wavelength were used to characterize the materials, enabling a rigorous exploration of their optical properties [72], [73], [74], [75], [76], [77], [78], [79], [80], [81], [82]. Maxwell's curl equation was used to determine the light absorption and carrier generation rates through finite difference time domain (FDTD) analysis [83] [84]. This type of FDTD analysis is often used for validating experimental works [85], [86]. For the input radiation source, AM 1.5G standard solar spectrum was used. Perfectly matched layer (PML) boundary conditions were employed at the top and bottom boundaries in the vertical (Y) direction, while periodic boundary conditions were applied along the horizontal (X) axis. In the FDTD analysis, the optical electric field distribution ($\overline{E_{op}}$) was calculated inside different layers. Using the optical electric field and the imaginary part of the complex dielectric constant, the absorbed power ($P_{abs}$) was calculated [87].

$$P_{abs} = -\frac{1}{2}\omega\left|\overrightarrow{E_{op}}(\vec{r},\omega)\right|^2 im\{\varepsilon(\vec{r},\omega)\},\qquad(1)$$

where $\omega$ is the angular frequency, $\overrightarrow{E_{op}}$ is the optical electric field distribution, and $im\{\varepsilon(\vec{r},\omega)\}$ is the imaginary part of the complex dielectric constant.

$$g(\vec{r},\omega) = -\frac{\pi}{h}\left|\overrightarrow{E_{op}}(\vec{r},\omega)\right|^2 im\{\varepsilon(\vec{r},\omega)\},\qquad(2)$$

where $g(\vec{r},\omega)$ is the electron-hole pair generation rate at position $\vec{r}$ and angular frequency $\omega$, and $h$ is Planck's constant.

$$G(\vec{r}) = \int g(\vec{r},\omega)d\omega,\qquad(3)$$

where $G(\vec{r})$ represents the total electron-hole pair generation rate at position $\vec{r}$ obtained by integrating $g(\vec{r},\omega)$ over all photon frequencies above the cutoff frequency, corresponding to the optical bandgap of the absorber material.

The electrical simulation follows the optical part, with the generation rate calculated from the optical simulation used as input. At first, the top cell and bottom cell electrical simulations were done separately. Poisson's equation, drift-diffusion equations, and continuity equations were used to find the electrical characteristic metrics of the top cell and bottom cell separately.

$$-\nabla\cdot(\varepsilon_{dc}\nabla V) = q\rho,\qquad(4)$$

where $\varepsilon_{dc}$ is the DC dielectric permittivity, $V$ is the electrostatic potential, and $\rho$ is the net charge density.

$$J_n = q\mu_n nE + qD_n\nabla n,\qquad(5)$$

where $J_n$ is electron current density, $\mu_n$ is electron mobility, $n$ is electron concentration, $E$ is the electric field, and $D_n$ is the diffusion coefficient for electrons.

$$J_p = q\mu_p pE - qD_p \nabla p, \tag{6}$$

where $J_p$ is hole current density, $\mu_p$ is hole mobility, $p$ is hole concentration, $E$ is the electric field, and $D_p$ is the diffusion coefficient for holes.

$$\frac{\partial n}{\partial t} = \frac{1}{q}\nabla \cdot J_n + G_n - R_n, \tag{7}$$

where $G_n$ is the electron generation rate, and $R_n$ is the electron recombination rate.

$$\frac{\partial p}{\partial t} = -\frac{1}{q}\nabla \cdot J_p + G_p - R_p, \tag{8}$$

where $G_p$ is the hole generation rate, and $R_p$ is the hole recombination rate.

The eqs. (4) to (8) were solved self-consistently to obtain the electrical characteristic metrics of the top and bottom cells, such as J-V characteristics, PCE, $V_{oc}$, $J_{sc}$, and FF. After performing the individual electrical simulation for the top cell and bottom cell separately, the tandem cell electrical simulation was performed. For the tandem cell simulation, the series circuit rules were applied.

$$V_{tandem} = V_{top} + V_{bottom}, \tag{9}$$

where $V_{top}$, $V_{bottom}$, and $V_{tandem}$ represent the applied bias voltages for the top, bottom, and tandem cells at the matched current of the tandem cell, respectively.

$$J_{tandem} = min(J_{top}, J_{bottom}), \tag{10}$$

where, $J_{top}$, $J_{bottom}$, and $J_{tandem}$ indicate the current densities for the top, bottom, and tandem cells, respectively. The PV parameters of the tandem cell, such as PCE, $J_{sc}$, $V_{oc}$, and FF, were calculated using the $V_{tandem}$ and $J_{tandem}$ values. The material data used in the electrical simulation, including doping concentration, carrier mobility, electron affinity, bandgap, recombination-related parameters, etc., are provided in the supplementary section S1.

## 3. Results and Discussion
## 3.1. Initial Thickness Optimization of CsSnGeI$_3$ and c-Si Layers

Initially, a sweep was performed on the absorber layers CsSnGeI$_3$ and c-Si. The thicknesses of the other layers were kept fixed, and the thicknesses of these two layers were varied. The radius of cylindrical Au nanorods was kept at 130 nm. The thicknesses and doping concentrations of different layers for this sweep are given in table 1.

**Table 1:** Initial thickness and doping concentration of different layers

| Layer | Initial thickness (nm) | Initial acceptor concentration (cm$^{-3}$) | Initial donor concentration (cm$^{-3}$) |
|---|---|---|---|
| ARC | 80 | - | - |
| FTO | 30 | - | 10$^{19}$ |
| TiO$_2$ | 30 | - | 10$^{18}$ |
| CsSnGeI$_3$ | - | 10$^{15}$ | - |
| Cu$_2$O | 30 | 10$^{18}$ | - |
| a-Si (n-doped) | 15 | - | 10$^{19}$ |
| GaSb | 150 | - | 10$^{16}$ |
| c-Si | - | - | 10$^{16}$ |
| a-Si (p-doped) | 15 | 10$^{19}$ | - |

| SiO$_2$ | 20 | - | - |
| Si$_3$N$_4$ | 100 | - | - |

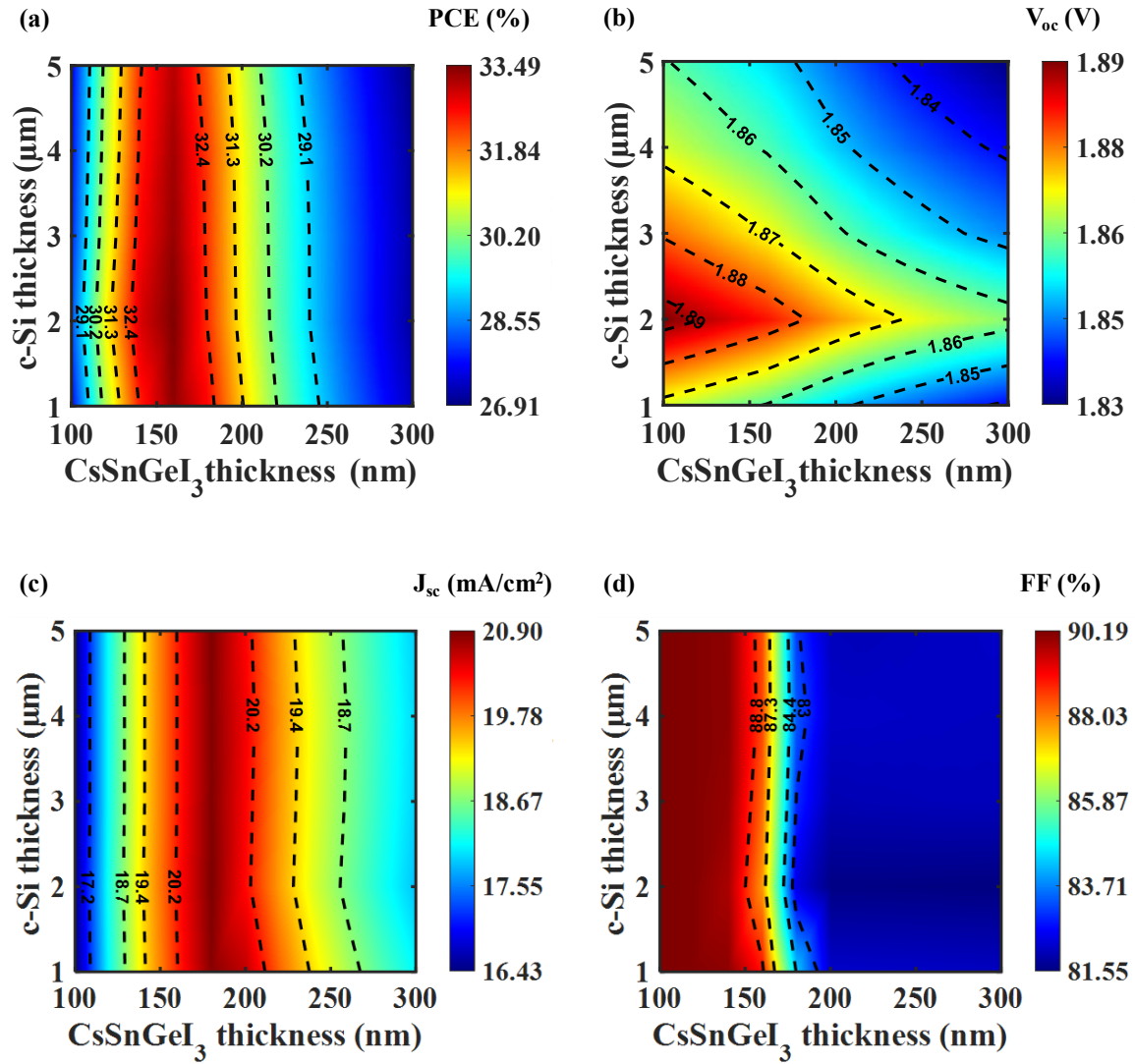

**Figure 3:** Variation of (a) PCE, (b) V$_{oc}$, (c) J$_{sc}$, and (d) FF of tandem cell as a function of CsSnGeI$_3$ and c-Si absorber layer thicknesses. Optimal performance occurs at a CsSnGeI$_3$ thickness of 160 nm and a c-Si thickness of 2 μm. The efficiency is mainly determined by the thickness of the CsSnGeI$_3$ layer, with only a slight effect from changes in the c-Si thickness.

Fig. 3 depicts the variation of key performance metrics that include PCE, V$_{oc}$, J$_{sc}$, and FF

as functions of the thicknesses of CsSnGeI$_3$ and c-Si absorber layers in a tandem solar cell setup. The peak PCE, illustrated in fig. 3(a), occurs when the CsSnGeI$_3$ layer is 160 nm thick and the c-Si layer measures 2 µm, achieving an efficiency of nearly 33.5%. PCE strongly depends on the thickness of the CsSnGeI$_3$ layer, but shows little sensitivity to changes in the c-Si thickness beyond 2 µm. As illustrated in fig. 3(b), the $V_{oc}$ of the tandem structure demonstrates an increase with the thickness of the c-Si layer up to 2 µm. Beyond this thickness, a slight decline is observed, indicating that the maximum $V_{oc}$ is attained at this particular thickness. Furthermore, the $V_{oc}$ progressively decreases with the increasing thickness of the perovskite layer, implying that a thinner perovskite layer is advantageous for achieving a higher $V_{oc}$. In fig. 3(c), $J_{sc}$ is mainly influenced by the CsSnGeI$_3$ layer, with peak values found in the 150-200 nm range. The $J_{sc}$ variation with c-Si thickness is minimal, indicating that the top absorber primarily controls series circuit current. In contrast, fig. 3(d) shows that FF remains fairly constant concerning c-Si thickness, but significantly declines as the CsSnGeI$_3$ layer thickens. A high FF, around 90%, is recorded when the CsSnGeI$_3$ layer is about 170 nm thick, while values fall below 85% when the thickness exceeds 200 nm. These results suggest that CsSnGeI$_3$ thickness is crucial for overall device performance, with c-Si thickness playing a limited role beyond 2 µm. Therefore, the optimal configuration for the tandem cell comprises a CsSnGeI$_3$ thickness of 160 nm in conjunction with a c-Si thickness of 2 µm.

## 3.2. Thickness Optimization of ETL and HTL Layers

The optimized thicknesses for perovskite and c-Si layers are 160 nm and 2 µm, respectively. Moreover, both a-Si layers have an optimum thickness of 15 nm (supplementary section S2). By applying these updated thicknesses for perovskite, c-Si, and a-Si layers, while maintaining the thickness of the other layers as before, the thickness variation was done for the ETL and HTL layers of the top cell of the tandem structure.

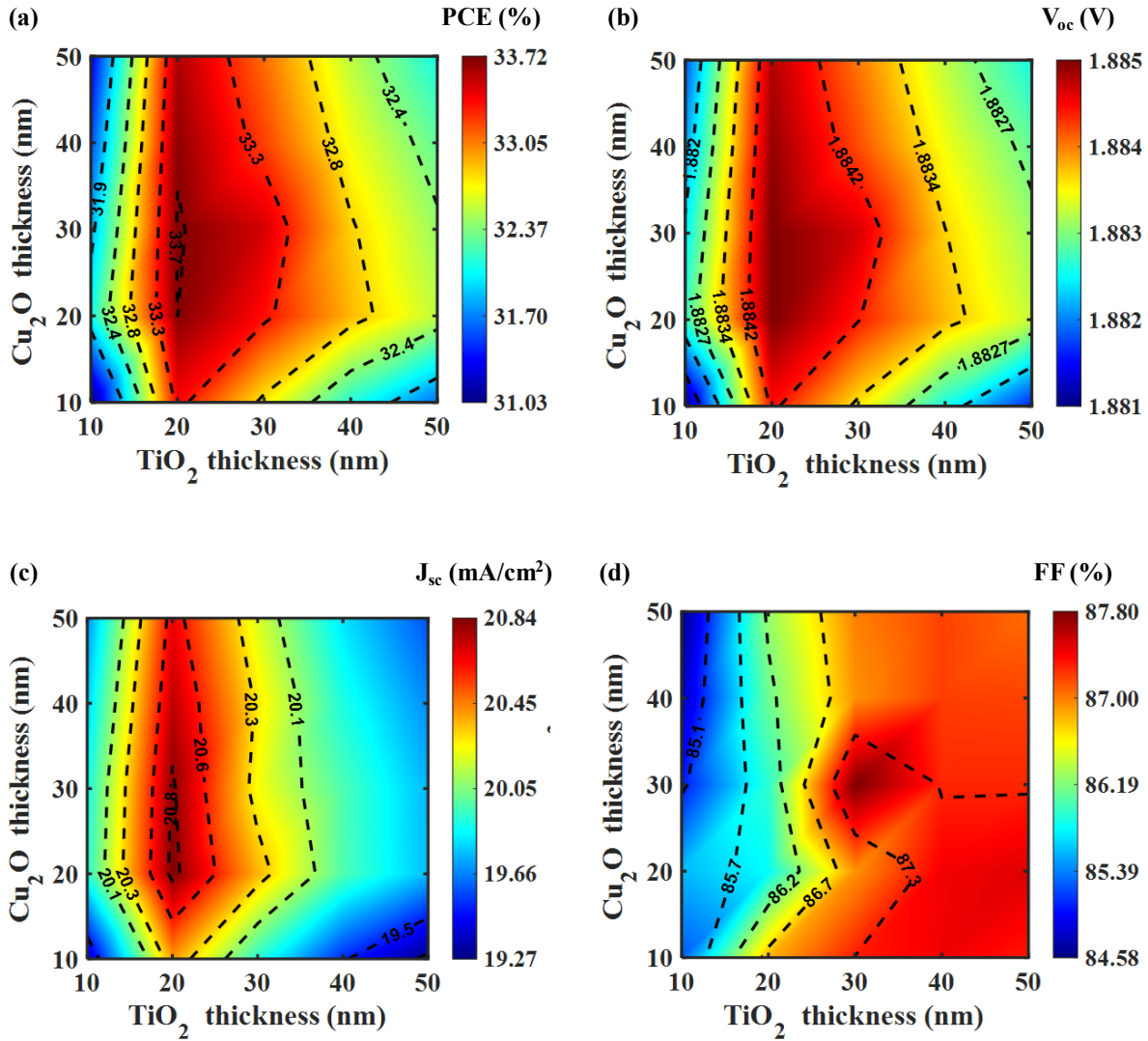

**Figure 4:** Variation of (a) PCE, (b) $V_{oc}$, (c) $J_{sc}$, and (d) FF of tandem cell as a function of $Cu_2O$ and $TiO_2$ transport layer thicknesses. Optimal performance is observed at a $TiO_2$ thickness of 20 nm and $Cu_2O$ thickness of 30 nm, corresponding to the highest PCE and $V_{oc}$. While the maximum $J_{sc}$ occurs at a $Cu_2O$ thickness of 20 nm, the highest FF is achieved when both layers are 30 nm thick.

Fig. 4 illustrates the influence of varying layer thicknesses of the $TiO_2$ ETL layer and $Cu_2O$ HTL layer on the performance metrics of the tandem solar cell. Fig. 4(a) shows that the optimum PCE is attained with a $TiO_2$ layer thickness of 20 nm and a $Cu_2O$ layer thickness of 30 nm. This configuration also yields the peak value for $V_{oc}$ as illustrated in fig. 4(b). However, fig. 4(c) shows a slight variation, revealing that the maximum $J_{sc}$ occurs with a 20 nm thick $Cu_2O$ layer instead of 30 nm, suggesting that thinner $Cu_2O$ layers boost current generation. In the case of FF, the highest value is achieved using 30 nm thickness for both layers, as seen in fig. 4(d). These results emphasize the complex interaction between the transport layers in tandem cells and the necessity to optimize each layer thickness separately to achieve maximum PV efficiency. In summary, the tandem cell reaches optimal overall performance with a $TiO_2$ thickness of 20 nm and a $Cu_2O$ thickness of 30 nm. These ultra-thin carrier transport layers enable efficient charge extraction, minimize recombination losses, and improve overall device performance. Then, the thickness optimization of GaSb, ARC, and rear passivation layers was done (supplementary sections S3, S4 and S5). The optimal thicknesses of GaSb, ARC, and rear passivation layers are 150 nm, 100 nm, and 30 nm, respectively.

## 3.3. Optimization of Plasmonic Nanorods

Optical simulation was employed to compare various dielectric materials, and the resulting $J_{ph\_top}$ (top cell photocurrent density) and $J_{ph\_bottom}$ (bottom cell photocurrent density) were utilized to identify a suitable dielectric medium. The thickness of the dielectric layer was kept at 100 nm.

**Table 2:** Photocurrent density data for various dielectric materials in the tandem structure

| Dielectric material | $J_{ph\_top}$ (mA/cm$^2$) | $J_{ph\_bottom}$ (mA/cm$^2$) |
|---|---|---|
| $MgF_2$ | 21.2388 | 22.4974 |
| $Si_3N_4$ | 21.2105 | 22.5704 |
| AZO | 21.2054 | 21.8701 |
| ZnO | 21.1697 | 22.3405 |
| $Al_2O_3$ | 21.1828 | 22.5374 |

Among the dielectric materials evaluated in Table 2, $Si_3N_4$ emerges as a particularly advantageous option. While it demonstrates competitive photocurrent densities in both the top and bottom cells, its broader significance resides in its established utility within Si-based PERC and SHJ solar cell architectures. Notably, its role in passivating layers within SHJ and PERC structures demonstrates its ability to reduce surface recombination losses, thus improving overall device efficiency [88]. Since the bottom cell in the tandem setup is Si-based, incorporating $Si_3N_4$ on the rear side enhances optical performance while also providing rear passivation. Therefore, $Si_3N_4$ stands out as the most advantageous dielectric medium for the nanorod-embedded plasmonic reflector layer in the proposed tandem device.

The Ag nanoparticles used in Jamil et al. [83] had a hemispherical shape with a 130 nm radius. In this study, cylindrical Au nanorods were used instead of hemispherical Ag nanoparticles. This change resulted in a slightly better photocurrent density in this tandem structure, as shown in table 3. $J_{ph\_top}$ and $J_{ph\_bottom}$ were calculated through optical simulation and were compared. Both nanoparticles had a 130 nm radius for the comparison in table 3.

**Table 3:** Comparison between nanoparticle material and shape

| Nanoparticle material | Nanoparticle shape | $J_{ph\_top}$ (mA/cm$^2$) | $J_{ph\_bottom}$ (mA/cm$^2$) |
|---|---|---|---|
| Ag | Hemispherical | 21.174 | 22.3254 |
| Au | Cylinder | 21.186 | 22.5704 |

Table 3 confirms that the cylindrical shape is optimal for the nanoparticles in this tandem structure. The optimal radius for these cylindrical nanorods is 150 nm (supplementary section S6).

The Au nanorods with a cylindrical shape used here are based on structures proven in experiments. For example, Park et al. successfully created high-aspect-ratio Au nanorods and verified their crystallinity and shape using Transmission Electron Microscopy (TEM) [59]. Kim et al. provided detailed TEM analyses of nanorods embedded in solar cells, emphasizing their size control and optical advantages [89]. Mendes et al. showed how plasmonic nanoparticles can be effectively embedded into dielectric layers to enhance rear-side light trapping in silicon solar cells [52]. These findings support the structural assumptions in our simulations and show that incorporating such nanostructures into practical devices is feasible.

## 3.4. Optimization of Doping Concentration

To enhance understanding of doping effects in various layers of the top and bottom cells, sweeping was conducted with different doping concentrations during electrical simulations.

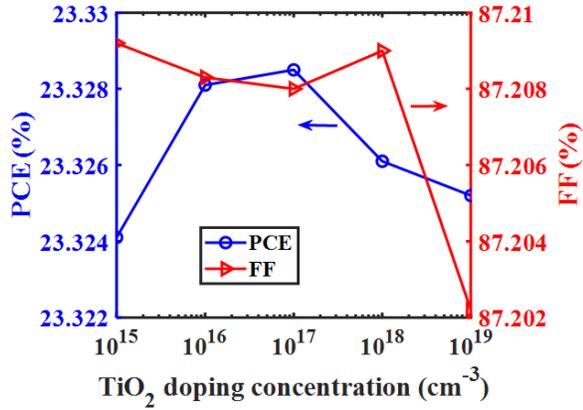
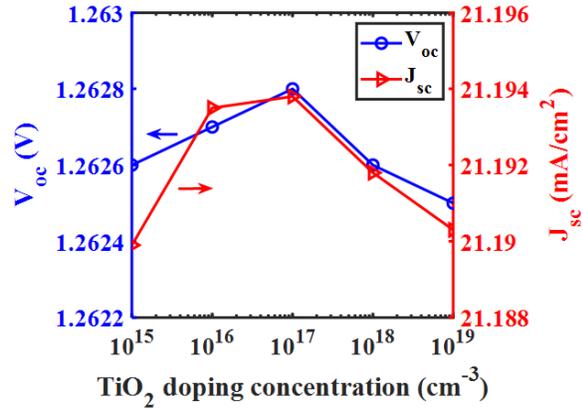

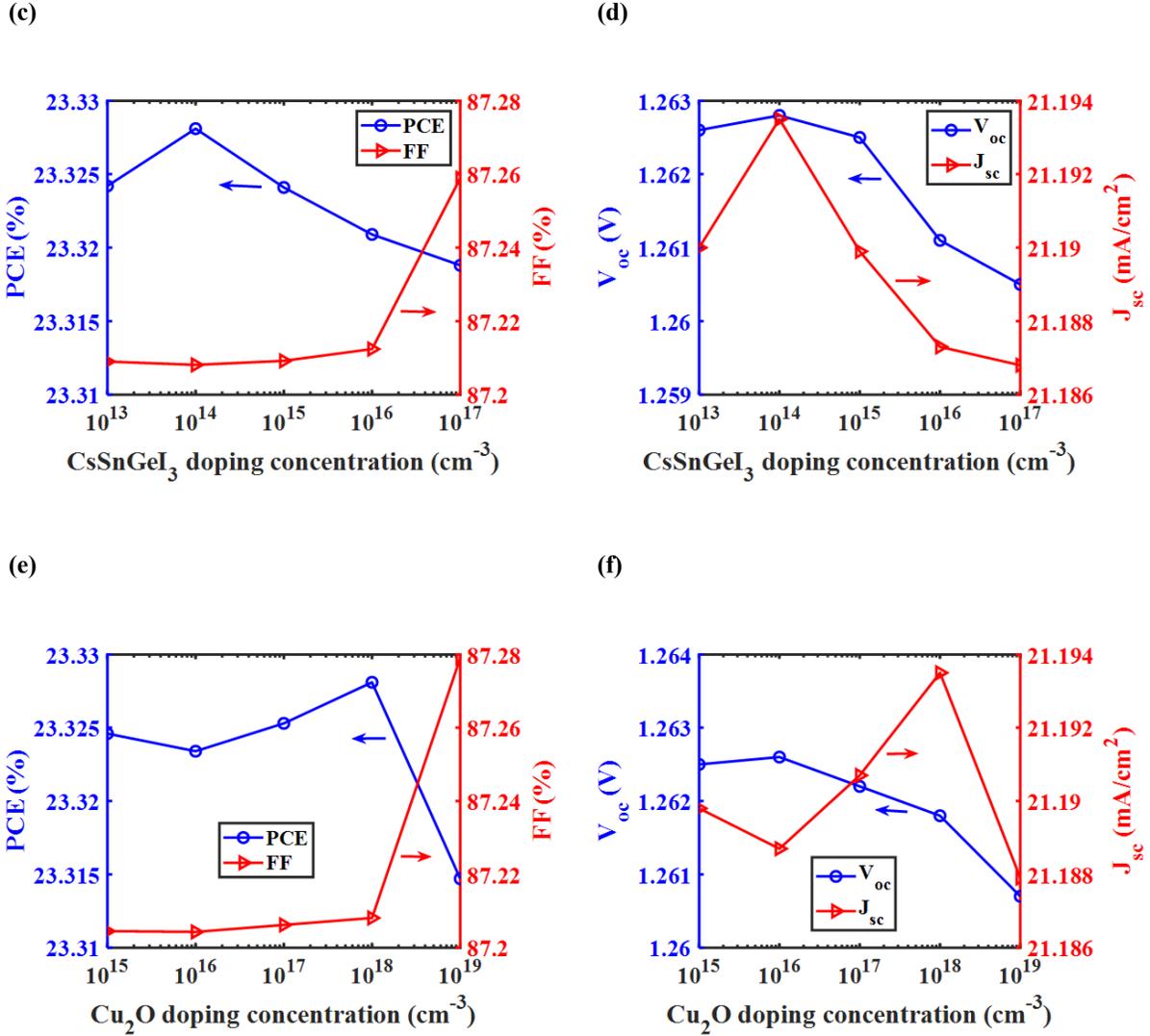

**Figure 5:** Variation in the PV performance of the top cell as a function of doping concentration for critical layers: (a) PCE and FF, and (b) $V_{oc}$ and $J_{sc}$ in the n-doped $TiO_2$ layer; (c) PCE and FF, and (d) $V_{oc}$ and $J_{sc}$ in the p-doped $CsSnGeI_3$ layer; (e) PCE and FF, and (f) $V_{oc}$ and $J_{sc}$ in the p-doped $Cu_2O$ layer.

Fig. 5(a) reflects the highest PCE achieved with an n-doped $TiO_2$ layer at a doping concentration of $10^{17}$ cm$^{-3}$. Figs. 5(a), and 5(b) reveal a consistent trend among them. The PCE, $V_{oc}$, and $J_{sc}$ values of the top cell rise with the ETL layer's doping concentration up to $10^{17}$ cm$^{-3}$, but start to decline if the concentration is further increased. Additionally, fig. 5(a) indicates that the top cell maintains a good FF value at this concentration. Fig. 5(c) illustrates that the highest PCE was achieved with a p-doped $CsSnGeI_3$ layer at a doping concentration of $10^{14}$ cm$^{-3}$. After this concentration, the PCE value started to decrease gradually. As shown in fig. 5(d), the $V_{oc}$ values were relatively consistent within the $10^{13}$ to $10^{15}$ cm$^{-3}$ range. In contrast, the $J_{sc}$ value in fig. 5(d) peaked at $10^{14}$ cm$^{-3}$, significantly outperforming the values observed at other concentrations. Even though the highest FF occurs at $10^{17}$ cm$^{-3}$ in fig. 5(c), the other parameters indicate that the optimal doping concentration for the perovskite absorber is $10^{14}$ cm$^{-3}$. Fig. 5(e) illustrates that the peak PCE occurs at a doping concentration of $10^{18}$ cm$^{-3}$ for the p-doped Cu$_2$O layer, showing a relatively high FF at this HTL layer doping concentration. Fig. 5(f) reveals that the $V_{oc}$ declines gradually after reaching a doping concentration of $10^{16}$ cm$^{-3}$. Nevertheless, the $J_{sc}$ is the highest at the $10^{18}$ cm$^{-3}$ doping concentration, establishing it as the optimal value. Thus, based on the trends observed in fig. 5, the optimized doping concentrations for TiO$_2$, CsSnGeI$_3$, and Cu$_2$O layers are $10^{17}$ cm$^{-3}$, $10^{14}$ cm$^{-3}$, and $10^{18}$ cm$^{-3}$, respectively. These optimized values of doping concentration are crucial in improving charge transport and enhancing the overall efficiency of the device.

(a)
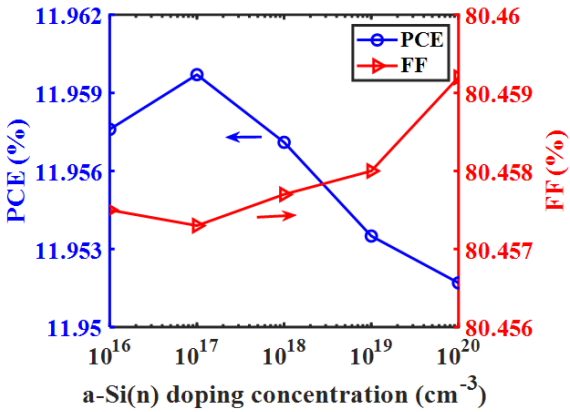

(b)
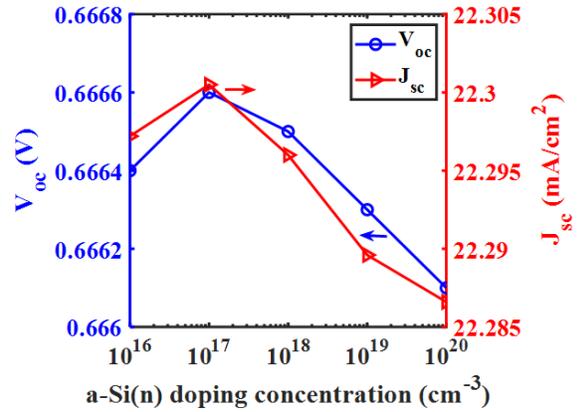

**(c)** 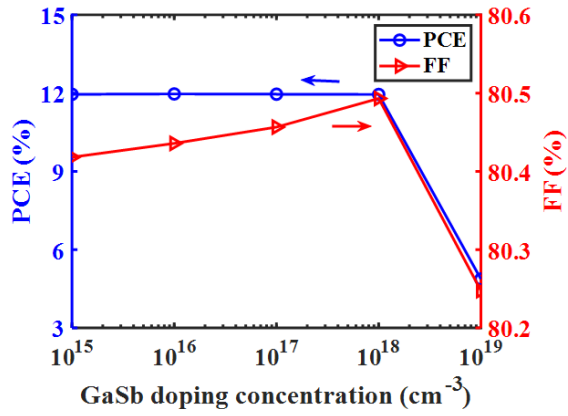

**(d)** 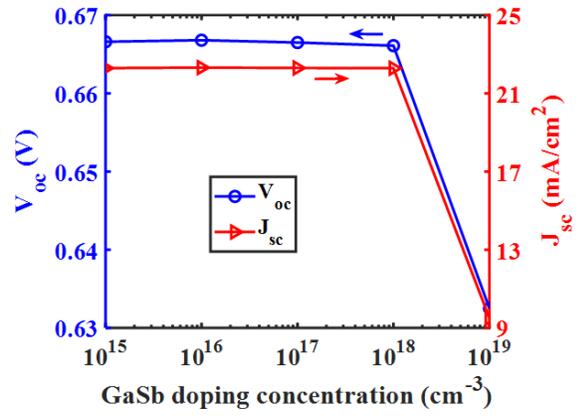

**(e)** 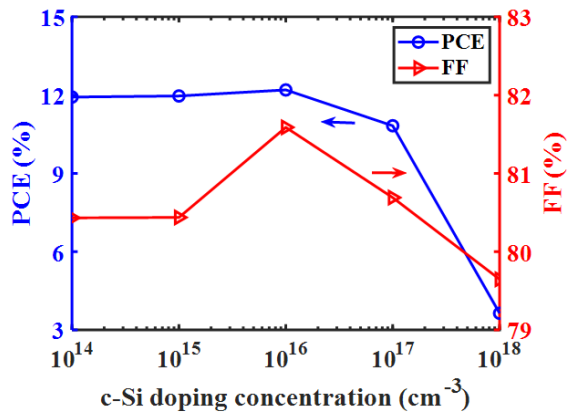

**(f)** 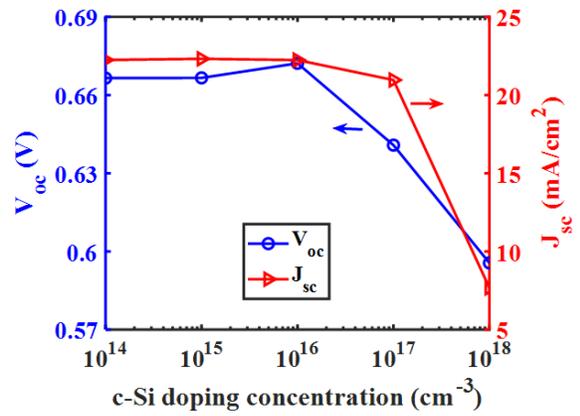

**(g)** 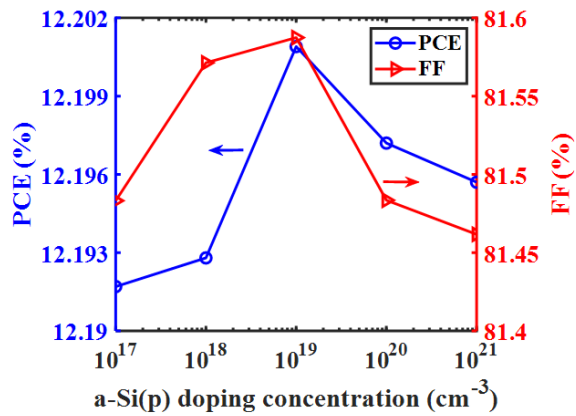

**(h)** 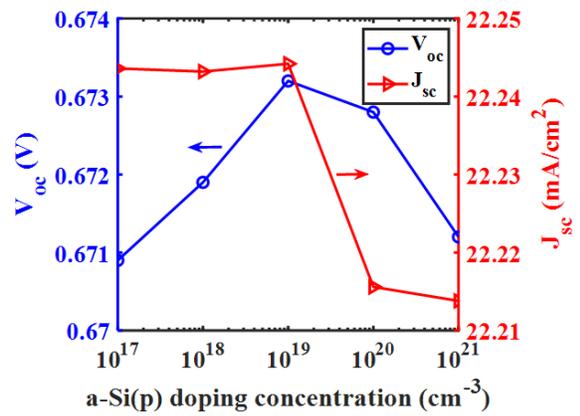

**Figure 6:** Variation in the PV performance of the bottom cell as a function of doping concentration for critical layers: (a) PCE and FF, and (b) $V_{oc}$ and $J_{sc}$ in the n-doped a-Si layer; (c) PCE and FF, and (d) $V_{oc}$ and $J_{sc}$ in the n-doped GaSb layer; (e) PCE and FF, and (f) $V_{oc}$ and $J_{sc}$ in the n-doped c-Si layer; (g) PCE and FF, and (h) $V_{oc}$ and $J_{sc}$ in the p-doped a-Si layer.

Fig. 6(a) demonstrates that peak PCE occurs when the n-doped a-Si layer has a doping concentration of $10^{17}$ cm$^{-3}$. This peak in PCE aligns with the observation that both $V_{oc}$ and $J_{sc}$ reach their maximum at the same concentration in fig. 6(b). Thus, the ideal doping concentration for the n-doped a-Si layer is $10^{17}$ cm$^{-3}$, even though its FF is the lowest at this level. Figs. 6(c) and 6(d) demonstrate that the PCE, $V_{oc}$, and $J_{sc}$ values remain almost constant across the doping concentration of the n-doped GaSb layer, which ranges from $10^{15}$ to $10^{18}$ cm$^{-3}$. The highest value is observed at a doping concentration of $10^{16}$ cm$^{-3}$ for all three metrics, indicating that this doping concentration is optimal for this layer, even though the FF is not at its maximum at this value. Fig. 6(e) shows that PCE and FF peak at a doping concentration of $10^{16}$ cm$^{-3}$ for the n-doped c-Si absorber layer. Additionally, fig. 6(f) illustrates that $J_{sc}$ remains nearly constant between $10^{14}$ and $10^{16}$ cm$^{-3}$, while $V_{oc}$ achieves its maximum at $10^{16}$ cm$^{-3}$, leading to the highest PCE at this concentration. Figs. 6(g) and 6(h) show that PCE, FF, $V_{oc}$, and $J_{sc}$ all peak at a doping concentration of $10^{19}$ cm$^{-3}$ for the p-doped a-Si layer. While $J_{sc}$ stays almost the same between $10^{17}$ and $10^{19}$ cm$^{-3}$, $V_{oc}$ reaches its highest point at $10^{19}$ cm$^{-3}$, leading to the maximum PCE at this concentration. Thus, from the trends depicted in fig. 6, the optimized doping concentrations for n-doped a-Si, GaSb, c-Si, and p-doped a-Si are $10^{17}$ cm$^{-3}$, $10^{16}$ cm$^{-3}$, $10^{16}$ cm$^{-3}$, and $10^{19}$ cm$^{-3}$, respectively. These precisely adjusted doping concentrations facilitate optimal overall device efficiency.

## 3.5. Final Thickness Optimization of CsSnGeI$_3$ and c-Si Layers

After optimization of all the layers through sweeping the thicknesses and doping levels, final thickness optimization for the CsSnGeI$_3$ and c-Si absorber layers was conducted. This step aimed to confirm if the prior optimizations, along with the thickness and doping variations of the various layers, affected these primary absorber layers of the tandem cell.

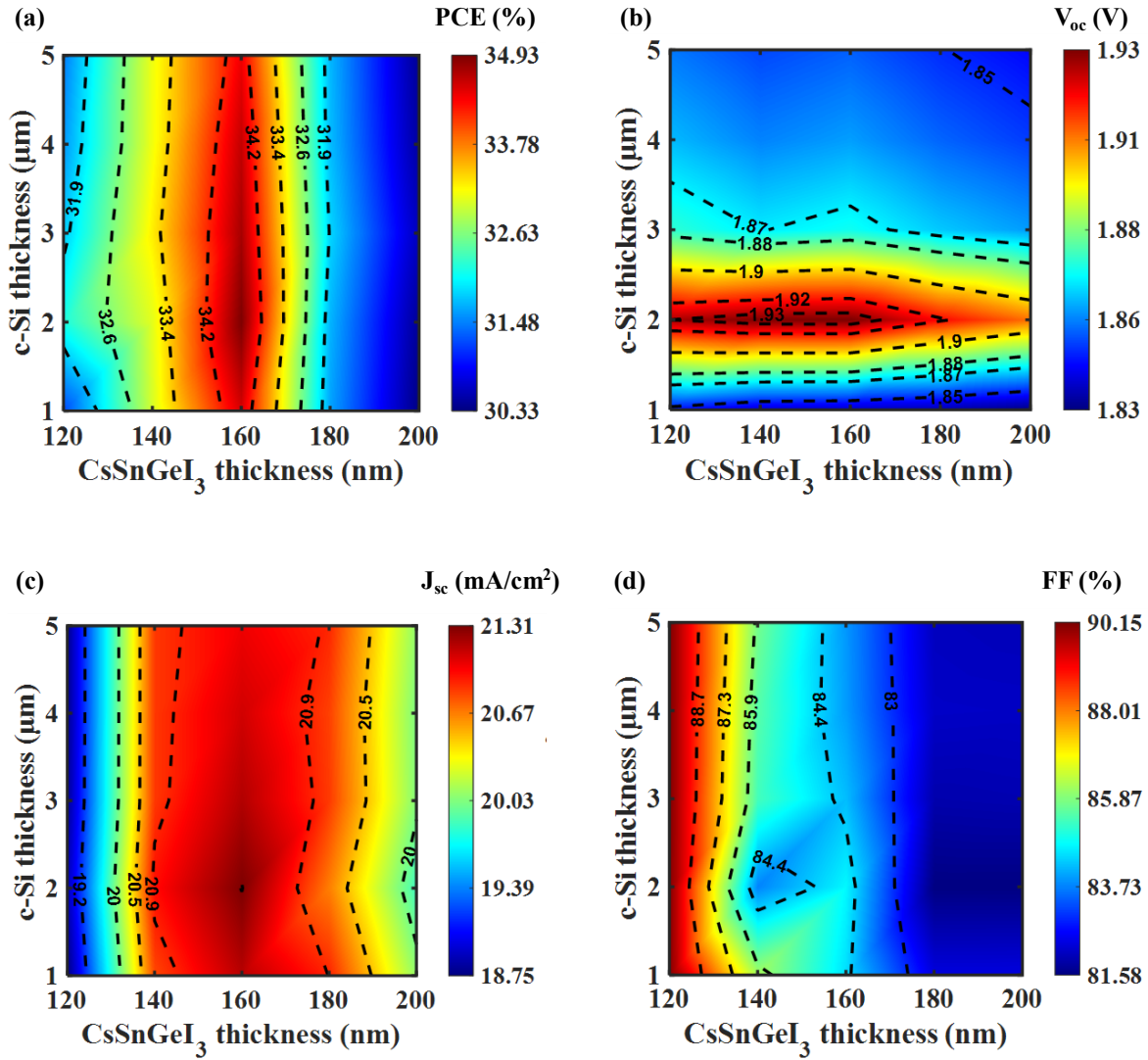

**Figure 7:** Variation of (a) PCE, (b) $V_{oc}$, (c) $J_{sc}$, and (d) FF of tandem cell as a function of thicknesses of CsSnGeI$_3$ and c-Si absorber layers. Optimal performance is achieved with a CsSnGeI$_3$ layer thickness of 160 nm and a c-Si layer thickness of 2 μm.

Fig. 7(a) demonstrates that the PCE peaks at a CsSnGeI$_3$ thickness of 160 nm, showing minimal variation across different c-Si thicknesses. In fig. 7(b), the maximum $V_{oc}$ occurs at a c-Si thickness of 2 μm, emphasizing its crucial role in influencing $V_{oc}$, while modifications in CsSnGeI$_3$ thickness result in only slight changes. Fig. 7(c) reveals that the highest $J_{sc}$ is observed at 160 nm for the CsSnGeI$_3$ layer, with only minor effects from variations in the c-Si layer thickness. Lastly, fig. 7(d) illustrates that the FF remains relatively high for the previously optimized thicknesses of both layers. These findings indicate that the ideal thicknesses for the CsSnGeI$_3$ and c-Si layers are 160 nm and 2 µm, respectively. This supports previous results, confirming that changes in thickness and doping concentrations across different layers do not affect the optimal thickness of these absorber layers.

## 3.6. Fully Optimized Tandem Cell Characteristic Metrics

After optimizing the tandem cell, the relationship between the current density and power density of the tandem cell as a function of the output voltage is demonstrated in fig. 8.

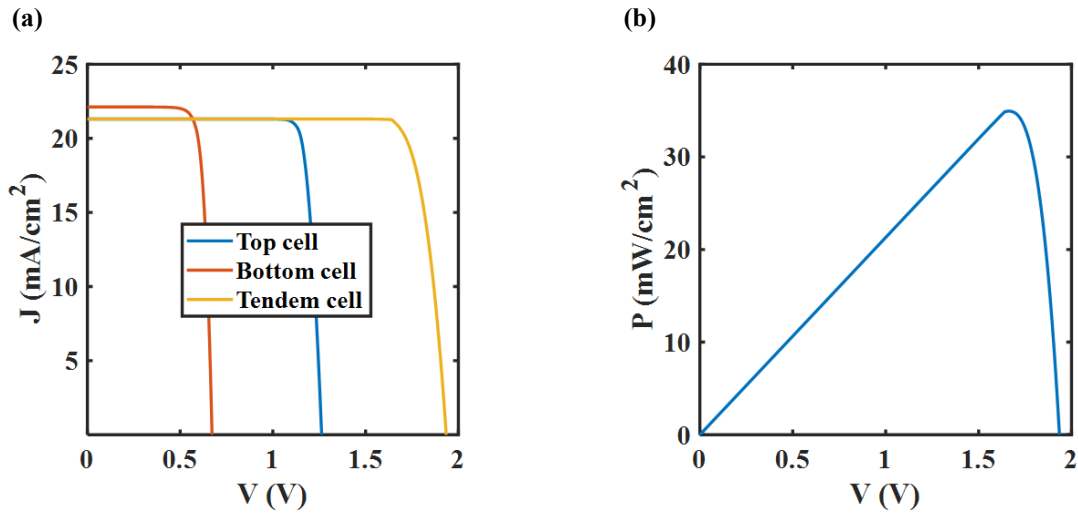

**Figure 8:** (a) Current density-voltage (J-V) and (b) power density-voltage (P-V) characteristics of fully optimized CsSnGeI$_3$/Si tandem cell.

Fig. 8(a) illustrates that the short circuit current density of the optimized tandem cell is constrained by that of the top cell. Fig. 8(b) shows that the tandem structure achieves maximum power at an applied voltage of 1.625 V. A band diagram illustrating the top and bottom cell structures at the Maximum Power Point Tracking (MPPT) conditions is provided in the supplementary section S7, offering a visual overview of charge carrier transport throughout the tandem device.

Under MPPT conditions, a 34.93% PCE is achieved, which exceeds the SQ limit. Additional PV characteristic metrics of the optimized structure are provided in table 4.

**Table 4**: Characteristic metrics of the final optimized structure under MPPT conditions

| Cell type | PCE (%) | $V_{oc}$ (V) | $J_{sc}$ (mA/cm$^2$) | FF (%) |
|---|---|---|---|---|
| Top cell | 23.46 | 1.26 | 21.30 | 87.18 |
| Bottom cell | 12.13 | 0.67 | 22.12 | 81.61 |
| Tandem cell | 34.93 | 1.93 | 21.30 | 84.74 |

Table 4 summarizes the characteristic metrics for the top, bottom, and tandem cells. The PCE of the tandem cell exceeds the PCE of the individual cells. The $V_{oc}$ of the tandem cell is particularly high, as it represents the combined $V_{oc}$ of both the top and bottom cells. However, the $J_{sc}$ of the tandem cell is constrained by the lower $J_{sc}$ of the two due to the series circuit rule of the tandem structure, which in this case is the $J_{sc}$ of the top cell.

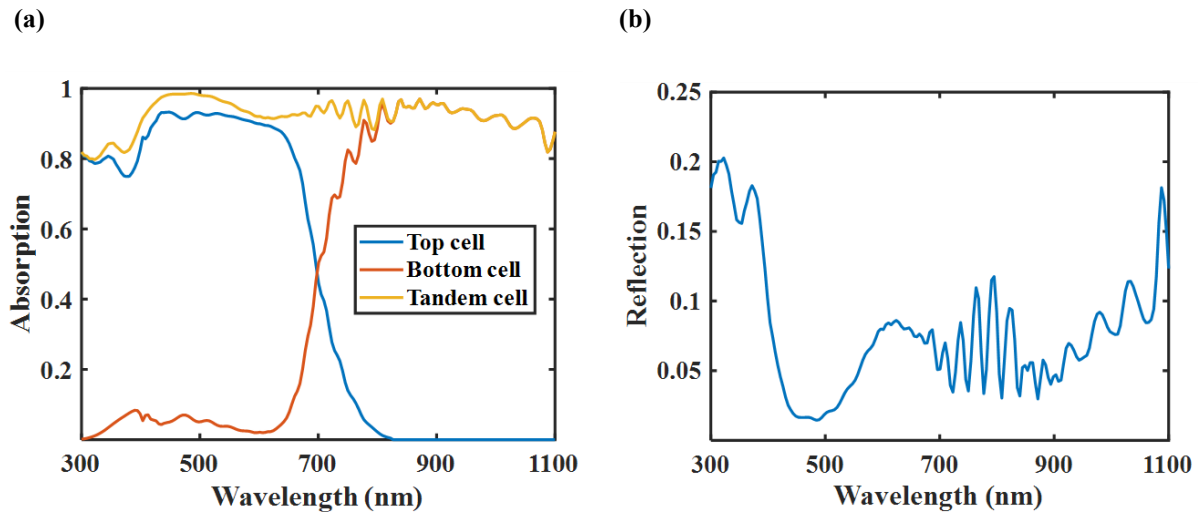

(c)

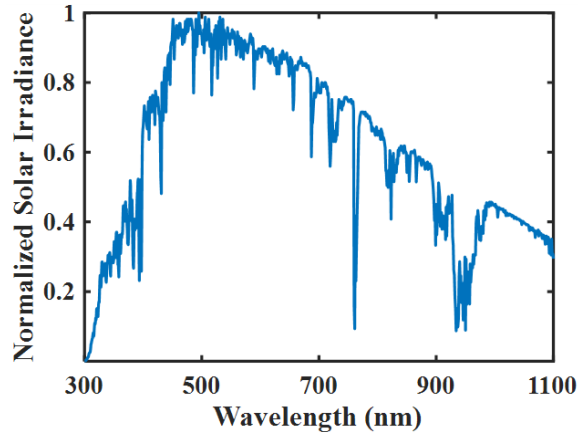

**Figure 9:** (a) Normalized absorption spectra of the top, bottom, and tandem cells, (b) normalized reflection spectra of the tandem cell, and (c) normalized AM 1.5G solar spectrum. The figures indicate that the top and bottom cells complement each other effectively, enabling the tandem configuration to achieve broad and enhanced absorption across the entire spectral range.

Fig. 9(a) illustrates the excellent absorption capabilities of the tandem cell, where the top and bottom cells work harmoniously across a broad spectrum of wavelengths. The absorption spectra indicate that the tandem structure effectively captures light throughout the range, with the top cell mainly absorbing shorter wavelengths and the bottom cell focusing on longer ones. This collaboration between the two cells significantly enhances the device's overall absorption, thereby boosting the efficiency of the tandem solar cell. Adding a thin GaSb layer to the SHJ bottom cell improved absorption, particularly in the NIR spectrum, and helped in reducing the thickness of the c-Si absorber layer. In fig. 9(b), the reflection spectra of the tandem cell show consistently low reflection across the entire spectral range, especially in the 450-600 nm region, where solar irradiance is the most intense. This low reflectivity is essential for maximizing light absorption instead of reflection, thus improving the device's overall performance. Fig. 9(c) presents the normalized AM 1.5G solar spectrum, illustrating the standard solar irradiance that reaches Earth's surface. The spectrum peaks in the 450-700 nm wavelength range, with the highest intensity near 550 nm as depicted in the figure. This wavelength range is essential for converting solar energy because it includes most of the energy available in sunlight. The absorption properties of the tandem cell, highlighted in fig. 9(a), align closely with this spectral region, indicating that the tandem configuration is effectively optimized for energy conversion from the solar spectrum.

**Table 5:** Comparison of characteristic metrics with other Si-based tandem cells

| Cell type | PCE (%) | $V_{oc}$ (V) | $J_{sc}$ (mA/cm$^2$) | FF (%) |
|---|---|---|---|---|
| This work | 34.93 | 1.93 | 21.30 | 84.74 |
| Perovskite/Si tandem cell [90] | 33.7 | 1.985 | 21.02 | 81.6 |
| Perovskite/Si tandem cell [46] | 32.2 | 1.87 | 20.3 | 84.7 |
| Perovskite/Si tandem cell [26] | 23.8 | 1.76 | 19.2 | 70 |
| Perovskite/Si tandem cell [48] | 28.2 | 1.88 | 19.6 | 78.6 |
| Kesterite/Si tandem cell [83] | 28.28 | 1.662 | 19.93 | 85.38 |
| AlGaAs/Si tandem cell [91] | 25.2 | 1.55 | 27.90 | 58 |

Table 5 shows that when combined with perovskites, Si-based tandem cells achieve higher PCE than other materials like kesterites. Generally, lead-based organic perovskites are associated with greater PCE values. Yet, this study reveals a significantly improved PCE using a lead-free inorganic perovskite, representing a significant advancement in perovskite-based tandem cell technology. This study achieves a favourable balance between $V_{oc}$ and $J_{sc}$, enhancing efficiency. In comparison, while the AlGaAs/Si tandem cell has a higher $J_{sc}$ [91], it demonstrates a significantly lower $V_{oc}$, resulting in a decreased PCE. Some of the perovskite/Si tandem cells shown in table 5 exhibit $V_{oc}$ levels comparable to those reported in this study [46], [48]. However, their $J_{sc}$ values are lacking, negatively affecting their overall performance. Ugur et al. fabricated a perovskite/Si tandem solar cell with a certified PCE of 33.7% that employs a lead-based organic-inorganic hybrid perovskite, surpassing the SQ limit [90]. In contrast, the present study demonstrates superior PV performance, including PCE, $V_{oc}$, $J_{sc}$, and FF, while employing a lead-free inorganic perovskite absorber.

Compared to earlier tandem designs, the CsSnGeI$_3$/SHJ architecture shows significant benefits. Although lead-based hybrid perovskite/Si tandems have achieved high efficiencies [26], [46], [48], their toxicity and environmental instability limit commercial viability. Conversely, fully inorganic, lead-free perovskites like MAGeI$_3$ [26], CsSnI$_3$ [42], and Bi or Cu-based options [28], [36] often face challenges such as lower carrier mobility, phase instability, or shorter diffusion lengths, resulting in modest tandem PCEs (~30-32%). CsSnGeI$_3$, however, features a suitable bandgap, better film quality, and increased environmental stability thanks to a native oxide layer [40], [43]. Its higher carrier mobility enhances charge transport and boosts current generation. The SHJ bottom cell enhances the top absorber by providing excellent surface passivation, high V$_{oc}$, and compatibility with well-established manufacturing techniques. This makes SHJ a more dependable choice compared to traditional homojunction or Passivated Emitter Rear Locally diffused (PERL) Si structures, or even less efficient bottom cells like CIGS or CZTS [6], [88]. To confirm the combined effectiveness of both absorbers, we simulated each subcell separately, achieving efficiencies of 23.46 % and 12.13%, as shown in Table 4. When combined in the optimized tandem design, the overall PCE increases significantly to 34.93 %, demonstrating that the tandem configuration and material synergy are crucial for exceeding the SQ limit. This highlights the vital contributions of the CsSnGeI$_3$ absorber and the SHJ structure in attaining high efficiency while maintaining environmental sustainability.

Recent advancements in perovskite/silicon tandem photovoltaics have demonstrated that achieving efficiencies in the mid-30% range is now possible at the laboratory scale. LONGi achieved a certified 34.85% efficiency for a two-terminal perovskite/Si tandem cell, and research institutions like Helmholtz-Zentrum Berlin (HZB) and King Abdullah University of Science and Technology (KAUST) have reported laboratory efficiencies between 32.5% and 33.7% [21], [92]. The simulated efficiency of 34.93% presented here slightly exceeds these results with a fully inorganic, lead-free perovskite, emphasizing its industrial relevance. The simulated efficiency of 34.93% presented in this work slightly exceeds these state-of-the-art values, demonstrating the performance potential of a fully inorganic, lead-free tandem design and underscoring its relevance for future high-efficiency PV architectures.

## 3.7. Impact of Plasmonic Nanorods

In order to assess the influence of plasmonic nanorods on device performance, simulations were executed utilizing the fully optimized tandem structure. For the purpose of comparative analysis, the cylindrical Au nanorods, along with their surrounding $Si_3N_4$ dielectric medium, were removed to model the structure without plasmonic enhancement. This approach facilitated a direct evaluation of the nanorods' impact on the device's optical behavior.

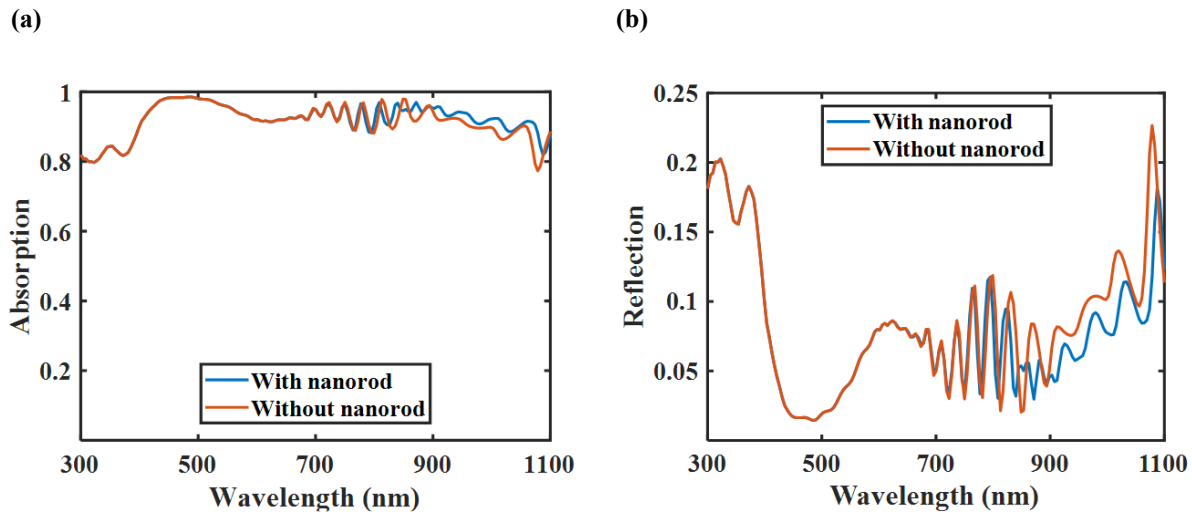

**Figure 10:** (a) Normalized absorption spectra of the tandem cell with and without nanorods, and (b) normalized reflection spectra of the tandem cell with and without nanorods. The incorporation of nanorods resulted in a slight increase in optical absorption, indicating their contribution to improved light harvesting.

Fig. 10 illustrates that the presence of plasmonic nanorods leads to a marginal enhancement in absorption and a slight decrease in reflection, especially at higher wavelengths. However, the overall spectral response remained consistently high across the visible spectrum in both situations, confirming that the tandem cell demonstrates strong light-harvesting capability even without plasmonic enhancement. This implies that while the nanorods provide additional light-trapping advantages, they are not mandatory for achieving high efficiency in the optimized configuration.

**Table 6:** Characteristic metrics of tandem structure with and without a plasmonic back reflector

| Cell type | PCE (%) | $V_{oc}$ (V) | $J_{sc}$ (mA/cm$^2$) | FF (%) |
|---|---|---|---|---|
| With a plasmonic back reflector | 34.93 | 1.9347 | 21.30 | 84.74 |
| Without a plasmonic back reflector | 34.32 | 1.9334 | 21.18 | 83.79 |

Table 6 shows that the PCE of the tandem structure drops from 34.93% to 34.32% when both the cylindrical Au nanorods and the dielectric layer on the rear side of the bottom cell are removed. This decrease is primarily attributed to a slight reduction in the $J_{sc}$, as the $V_{oc}$ remains nearly constant despite the removal of the plasmonic back reflector. This finding differs from the research by Jamil et al. [83], where the elimination of the plasmonic back reflector resulted in a PCE drop of more than 7%. In contrast, this study reports a decline of only 0.61%, suggesting that the structure could still exceed the SQ limit, even without the plasmonic back reflector. This is accomplished by utilizing an appropriate perovskite absorber in the top cell and incorporating an additional ultra-thin GaSb absorber layer within the bottom SHJ cell. Thus, if minimizing complexity and production costs is a priority, the plasmonic back reflector and the dielectric layer can be omitted from the design while maintaining a high efficiency.

## 3.8. Limitations and Future Outlook

This study offers a detailed numerical simulation of a lead-free perovskite/SHJ tandem solar cell with plasmonic enhancement. However, it is essential to note that the findings are based solely on FDTD and drift-diffusion models. While the study carefully considered realistic material properties, interfacial recombination (through surface recombination velocity), and fabrication practicality, no experimental validation has been conducted yet. Furthermore, issues such as the quality of interface passivation, long-term material stability, and challenges in layer deposition may influence actual performance. Future efforts will involve collaborating with experimental groups to develop and confirm the proposed design experimentally.

## 4. Conclusion

In summary, this study presents a fully inorganic and lead-free CsSnGeI$_3$/SHJ tandem solar cell that is simulated to achieve a remarkable PCE of 34.93%, surpassing the SQ limit for single-junction cells. The integration of CsSnGeI$_3$, a non-toxic and inorganic perovskite, directly addresses the pressing environmental concerns associated with traditional lead-based counterparts. The inclusion of a GaSb auxiliary absorber within the SHJ bottom cell enhances optical absorption, particularly in the NIR region, allowing for the utilization of a thinner c-Si layer without compromising performance. Performance is additionally enhanced by incorporating cylindrical Au nanorods within a Si$_3$N$_4$ dielectric medium positioned at the rear of the bottom cell, thus amplifying light absorption through plasmonic effects. Notably, the structure retains high efficiency even in the absence of the plasmonic back reflector, providing flexibility in fabrication and cost reduction. The design illustrates a pragmatic trade-off between complexity and performance, paving the way for low-toxicity and cost-effective photovoltaic technologies. These findings are based on comprehensive numerical modeling and await experimental validation.

Opportunities for future work include enhancing this architecture through the application of surface texturing techniques, particularly to the top cell, to further improve light trapping and reduce reflection as typically observed in PERC cells. Incorporating additional junctions with suitably engineered bandgap materials may enable multi-junction configurations that capture a broader portion of the solar spectrum, thereby further enhancing the efficiency. These avenues, coupled with ongoing advancements in lead-free materials, establish the groundwork for next-generation high-efficiency tandem PV cells.

## Data Availability

Data that supports the findings of this work is available from the corresponding author upon reasonable request.

# Declaration of Interest

The authors have no conflict of interest to declare.

# Acknowledgements

Ehsanur Rahman acknowledges a basic research grant from BUET.